\documentclass[final,3p,times,twocolumn]{elsarticle}

\usepackage{amssymb}
\usepackage{xcolor}
\usepackage{color,soul}
 \usepackage{amsmath}
\usepackage{rotating}
\usepackage{siunitx}
\usepackage{hyperref}
\usepackage{booktabs}

\journal{Computational}

\begin{document}

\begin{frontmatter}



\title{A ReaxFF-based thermomechanical analysis of N-carbophenes: phase-change, thermal expansion, and high temperature synthesis pathway}

\author[label1]{Chad E. Junkermeier\corref{cor1}}
\author[label1]{Kat Lavarez}
\author[label1,label2]{R. Martin Adra}
\author[label1]{Valeria Aparicio Diaz}
\author[label3, label4]{Heather Osterstock}
\author[label5,label6]{Pal Casinto}
\author[label5,label7]{M. Verano}
\author[label8]{Ricardo Paupitz}
\author[label9]{Adri C. T. van Duin}

\address[label1]{Department of Physics and Astronomy, University of Hawai`i at Mānoa, Honolulu HI, USA}
\address[label2]{Present address: Phil and Penny Knight Campus for Accelerating Scientific Impact, Eugene OR, USA}

\address[label3]{Department of Biochemistry, University of Washington, Seattle WA, USA}
\address[label4]{Present address: Dotquant LLC; Seattle WA, USA}

\address[label5]{Department of Science, Technology, Engineering, and Mathematics, University of Hawai`i Maui College, Kahului HI, USA}
\address[label6]{Present address: Department of Electrical and Computer Engineering, University of Hawai`i at Mānoa, Honolulu HI, USA}
\address[label7]{Present address: Department of Physics, New Mexico State University, Las Cruces NM, USA}

\address[label8]{Departamento de F\'{\i}sica, IGCE, Universidade Estadual Paulista, UNESP, 13506-900, Rio Claro, SP, Brazil}
\address[label9]{Department of Mechanical Engineering, Pennsylvania State University, University Park PA, USA}
\cortext[cor1]{junkerme@hawaii.edu}

\begin{abstract}

N-carbophenes are a class of two-dimensional covalent organic frameworks with potential for solid-state gas storage and as 2D topological materials. Previous studies have demonstrated that variations in their bonding, topology, and functionalization enable the tuning of their chemical, electrical, and mechanical properties.  Yet, the thermal stability and high-temperature behavior of pristine and functionalized N-carbophenes remain unexplored. Using ReaxFF-based reactive molecular dynamics (RMD) simulations with extensive statistical validation, we performed temperature-ramp MD simulations of pristine and functionalized N-carbophenes. We demonstrate that N-carbophenes remain stable up to temperatures above $1000$ K. The phase-change onset temperatures decrease as the N-phenylene chain length increases in pristine N-carbophenes, attributed to increasing antiaromaticity in the central phenylene segments, thereby contributing to the foundational understanding of aromatic versus antiaromatic bonding in 2D carbon networks, a topic of considerable interest in theoretical chemistry. Pristine N-carbophenes exhibit negative area thermal expansion (NATE), whereas functional groups modulate this, leading to either negative or positive expansion. Functional groups remain stably bonded well above the transition temperature. We also show that a temperature-induced phase transition from graphenylene (2-carbophene) to $\gamma$-graphyne is possible. Our results provide upper bounds on N-carbophene stability, clarify the relationships between structure and thermal properties, and identify a new transformation pathway. These results will have applications in tunable band gaps, porous architectures, or chemically accessible sites.

\end{abstract}


\end{frontmatter}


\section{Introduction}

Two-dimensional covalent organic frameworks (2DCOFs) have been extensively studied because changes in their network structures often result in significant alterations in their electronic, mechanical, and chemical properties~\cite{Cote20051166,Wang2022202102290, Zhang20221c00693, ni2022engineering, fu2025fundamentals}. Similarly, the synthesis of novel 2D allotropes has spurred research into exciting applications requiring tunable band gaps, porous architectures, or chemically accessible sites~\cite{Ding5b10754, Mitra7b00925, Wang7b02648, C7CE00344G, Raju20183969}.

N-carbophenes—a class of biphenylene-based 2DCOFs—represent one such emerging class of materials~\cite{du1740796, junkermeier2019simplecarbophene, Junkermeier2022Covalent}. The biphenylene-based topology of 4- and 6-member rings surrounds large intrinsic pores with reactive edge sites. Previous work on N-carbophenes has demonstrated that they exhibit favorable formation energies, open-pore stacking configurations, and higher-order topological insulating behavior~\cite{junkermeier2019simplecarbophene, Yang2025033101}. Both direct calculation and quantitative structure-property relationships predictions indicate that N-carbophenes' elastic moduli decrease as N increases, with strain-induced band gap openings by as much $80\%$ for uniaxial strains and $160 \% $ for biaxial strains~\cite{BATISTA2023112103, arockiaraj2024topological}. Functionalizing pristine N-carbophenes lowers their formation energies, with some cases yielding negative formation energies~\cite{Junkermeier2022Covalent}. It has also been demonstrated that functionalized N-carbophenes may be used in gas capture and storage applications~\cite{JUNKERMEIER2024112665, JUNKERMEIER2024110652}. Further, due to their strain-tunable band gaps and 2nd order topological insulating properties, N-carbophenes may be useful in next-generation semiconductor device engineering. Finally, due to their stress-strain relationships and porous nature, they may be useful in applications like molecular fishnets~\cite{Servalli2018fishnet}.

Despite the exciting physical, chemical, and electrical properties of N-carbophenes, their thermal behavior remains poorly understood. No detailed study of thermal stability exists, and the effects of N-phenylene chain length or functionalization on high-temperature behavior are also unknown. Furthermore, no systematic evaluation of temperature-induced phase transformations has been conducted.

\section{Methodology}\label{sec:methods}

This project was initiated at the request of the Chancellor of the University of Hawai‘i Maui College, with the goal of translating our research into an accessible outreach initiative for high school students. Its primary objective was to introduce students to fundamental concepts in materials science and to cultivate their curiosity about scientific discovery. Accordingly, rather than automating the workflow, we incorporated student participation in the analysis of molecular dynamics (MD) simulations.

Terminology used throughout this work is summarized in Table~SI.1 of the Supporting Information.


The ReaxFF reactive force field method, as implemented in the stand-alone SCM Amsterdam Modeling Suite 2019.3, was used to determine N-carbophenes’ phase-change onset as temperature increases~\cite{senftle2016reaxff, scmReaxFF2019}. Model crystal structures were relaxed using conjugate gradient energy minimization with a 0.5 kcal/mol convergence criterion, allowing all cell parameters to change. Relaxed structures underwent a 600,000 time step (0.025 fs/time-step) MD simulation at 10 K in an isothermal-isobaric (NPT) ensemble with Anderson-Hoover barostat and thermostat (denoted AHNPT in ReaxFF)~\cite{martyna1996explicit}. During a constant-temperature MD simulation, a checkpoint file (a molsav file in ReaxFF) containing atomic positions and velocities was written every 10,000 time steps. Thus, for each N-carbophene model studied, a relaxed structure, a constant-temperature MD simulation, and 60 molsav files were generated. The 11th through 60th molsav files were used as initial conditions for temperature ramp MD simulations. Using the same ensemble and thermostat as in the constant-temperature simulation, each temperature-ramp MD simulation linearly increased the system’s temperature from 10 K to 2010 K over 2 million time steps (0.001 K/time-step, 0.25 fs/time-step). Thus, each N-carbophene structure underwent 50 independent temperature-ramp MD simulations. Replicating the temperature-ramp MD simulations 50 times enabled us to estimate variability and statistical power better. A bond-order cutoff of 0.001 was used for valence and torsion angles; EEM was applied over the system, and charge energy was used in total-energy calculations~\cite{Mortier19864315}. The simulations were produced using Kowalik \textit{et al.’s} reactive force field parameterization~\cite{kowalik2019atomistic, Gao2020eaaz4191, MAO202025}.

\begin{figure}
    \centering
    \includegraphics[width=\linewidth]{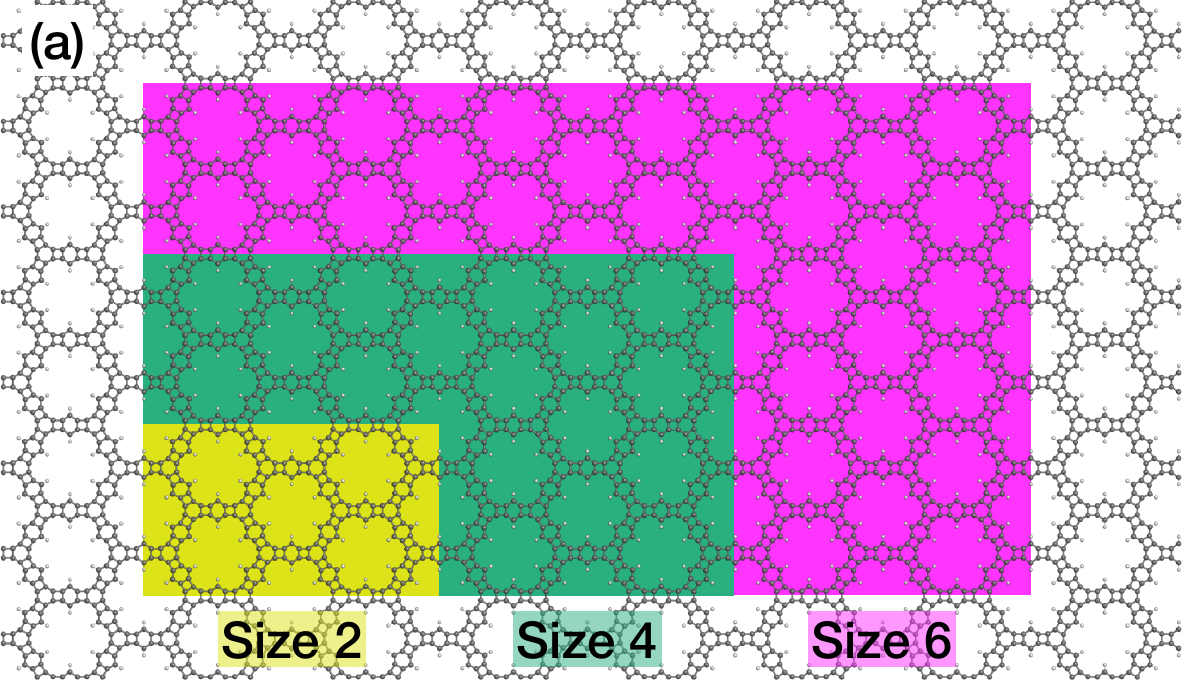}
    \includegraphics[width=\linewidth]{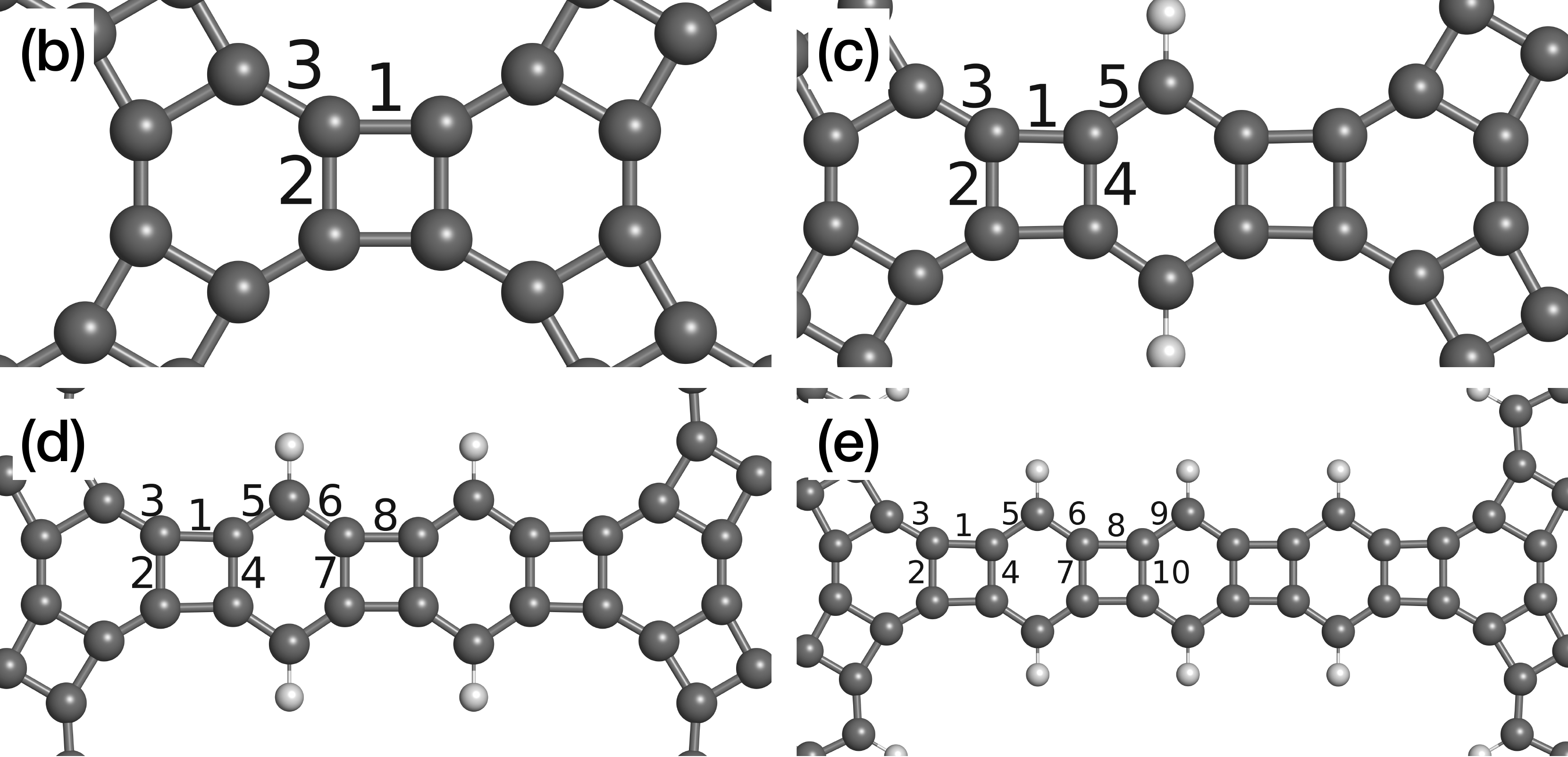}
    \caption{(a) A 3-Carbophene model that showcases different supercell sizes: size 2, 4, and 6. Nomenclature of bond types in (b) graphenylene, (c) 3-carbophene, (d) 4-carbophene, and (e) 5-carbophene~\cite{PyMOL}.}
    \label{fig:3cSupercellSizes}
\end{figure}

MD simulations were run on 17 crystal models that differed in N-carbophene type, supercell size, and attached functional group. Figure~\ref{fig:3cSupercellSizes} (a) shows three ground state 3-carbophene supercell models (Size 2, Size 4, Size 6). Examples of the graphenylene, 4-carbophene, and 5-carbophene supercells are in SI.Figure~1 of the Supporting Information. The 17 MD simulation groups are named by concatenating N-carbophene type (graphenylene = ``2carb", 3-carbophene = ``3carb", 4-carbophene = ``4carb", 5-carbophene = ``5carb"), model size (e.g., Size 2 = ``2"), and functional group (pristine = ``P", amine = ``NH2", carbonyl = ``CO", carboxyl = ``COOH", hydroxyl = ``OH", nitro = ``NO"). For example, the pristine Size-6 4-carbophene is 4carb6P, and the Size 2 amine-functionalized 3-carbophene is 3carb2NH2. For visualization, a flowchart of the MD simulations for one crystal model (e.g., 3carb2NH2) is in SI.Figure~2 of the Supporting Information.


We define the onset of a temperature-induced phase-change in each replicate as the temperature at which the first inter-atomic bond breaks. For each replicate, we determined this temperature by visual inspection in VMD (Visual Molecular Dynamics). We used the automatic pictorial linking and unlinking of bonded and non-bonded atoms controlled by the DynamicBonds representation, with a bond cutoff of 1.85 \AA~\cite{HUMPHREY199633, ISHIGAKI2018795}.  

A process was implemented to ensure that each replicate was analyzed multiple times by students. For each analysis, students recorded several pieces of information: the structure name (which included metadata about the simulation); the temperature at which the first bond broke; the type of bond that broke (Figures~\ref{fig:3cSupercellSizes} (b)-(e) provide bond type nomenclature); and whether the broken bond crossed a periodic boundary.

Student variability in recorded values of each replicate was expected, given limited prior training. Consequently, variation in the recorded temperatures resulted in a squared uncertainty of $0.072$. Additionally, MD results were recorded every 100 time steps, giving an uncertainty of $0.1$ K. Further, if the maximum difference in the recorded temperatures for a particular replicate was less than or equal to 0.2 K, the minimum value reported for that replicate was used. These uncertainties were then combined to give a total temperature measurement uncertainty of $0.35$ K. When analysis results differed by more than 0.2 K, the simulation was referred to an expert, who reviewed the MD simulation and determined the correct result by resolving discrepancies. Finally, extreme temperature outliers were removed using the interquartile range (IQR) method~\cite{rstatix}.

The Shapiro-Wilk test indicated that the temperature data were not normally distributed in five of the crystal models, necessitating a non-parametric statistical analysis~\cite{tseries}. The Wald-Wolfowitz Runs Tests suggested that the original temperature data were not randomly distributed for the crystal model 3carb2COOH~\cite{runstest}. We reran the constant-temperature MD simulation for 3carb2COOH to increase the number of time steps between the temperature ramp simulations. More temperature ramp simulations were run and analyzed, and the resulting distributions exhibited a random pattern. Potential periodic boundary condition errors were evaluated using a chi-squared test, which compared the expected and observed bond breaks across boundaries. This test showed no significant boundary-related artifacts~\cite{Patefield198191}. Overall, the statistical analyses support the validity and reliability of the MD simulation results.

\section{Results and Discussion}

\subsection{Comparison with experiment} 

\begin{figure}
    \centering
    \includegraphics[clip,width=3 in, keepaspectratio]{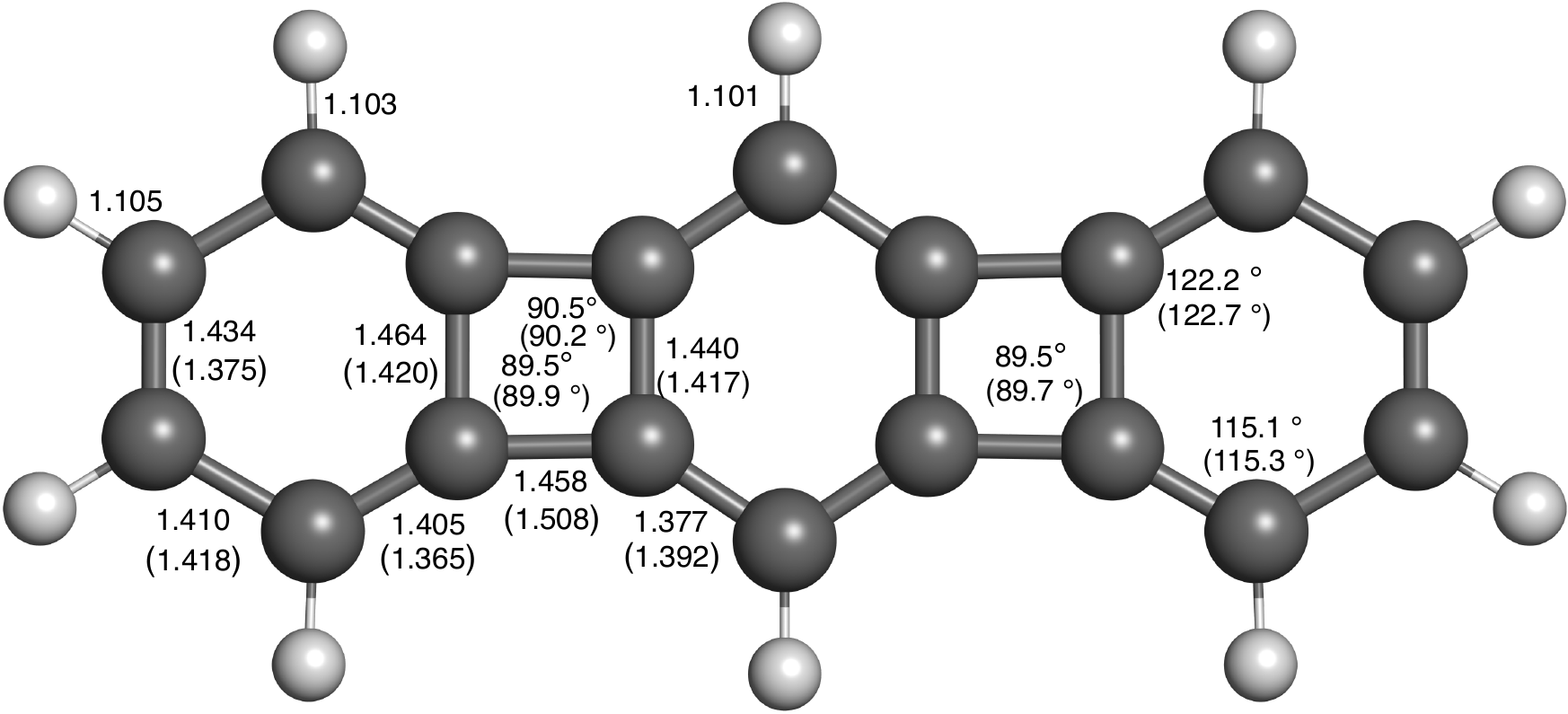}
    \caption{Comparison of the bond lengths in Angstroms and valence bond angles in degrees of 3-phenylene as computed by ReaxFF and experimentally determined through XRD (parenthesized values).}
    \label{fig:3phenylene_bond_angle}
\end{figure}

Since bond-length analysis is used to determine the phase-change onset, a comparison of the computed bond lengths of a 3-phenylene molecule with their experimental values was conducted to validate the accuracy of the subsequent results. Figure~\ref{fig:3phenylene_bond_angle} presents selected bond lengths and valence bonds calculated using ReaxFF, alongside the experimental values obtained from X-ray diffraction (XRD) and other sources~\cite{SCHLEIFENBAUM20017329}. The computed values closely match the empirical values, with a deviation of less than $4.3\%$. Notably, the experimentally measured out-of-plane dihedral angles are less than $0.7\%$, which aligns with the planar structure reported in prior research~\cite{SCHLEIFENBAUM20017329}. While the calculated C-H bond lengths are presented, the experimental C-H bond lengths are not included due to the limitations of XRD in accurately determining the positions of hydrogen atoms.

\subsection{Phase-change onset} 

\begin{figure*}
    \centering
    \includegraphics[clip,width=\linewidth, keepaspectratio]{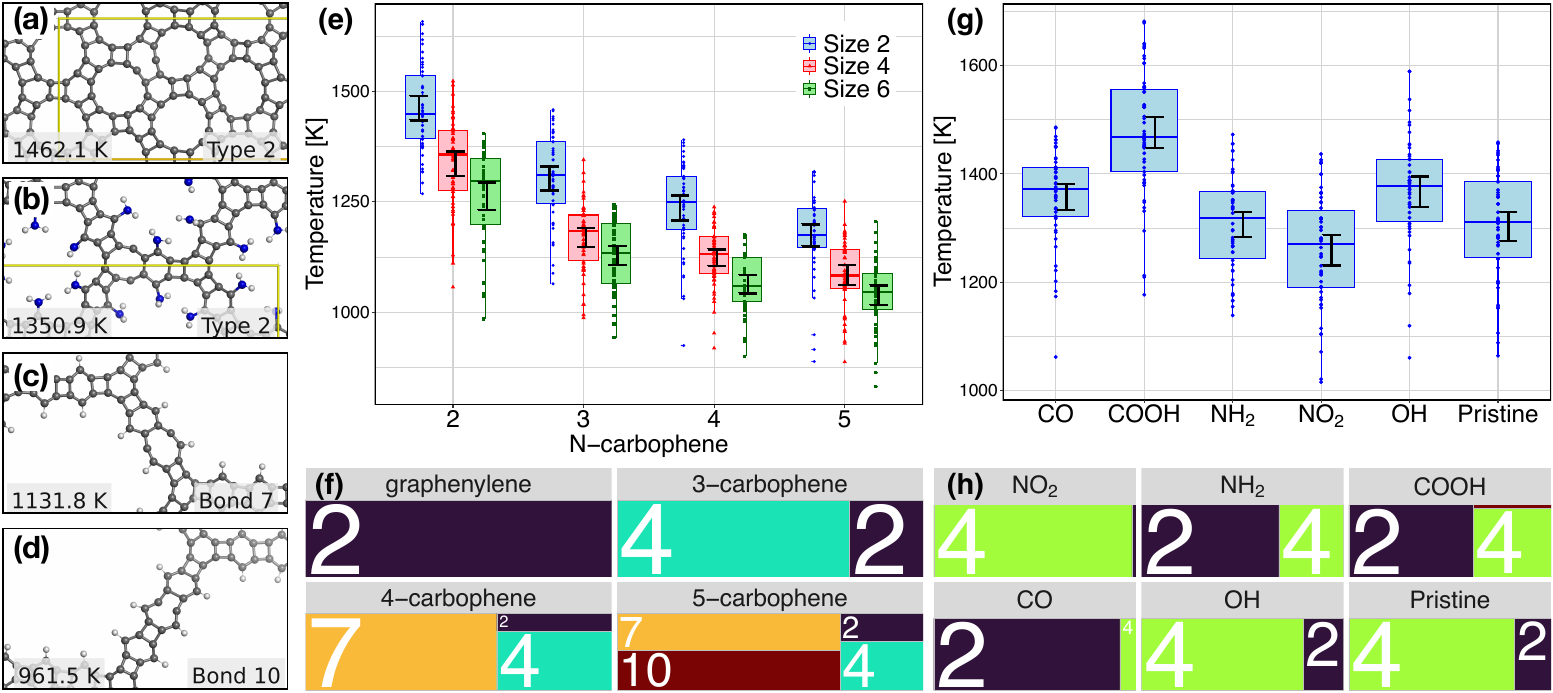}
    \caption{(a) - (d) Examples of the first bond breaking in each type of material in this study~\cite{patchwork}. Each frame presents the type of bond that broke and the temperature at which it broke. Gold lines outline supercell boundaries. Initial bond breaking temperatures of each type of (e) pristine N-carbophene (g) functionalized 3-carbophene. Treemaps of the ratios of the type of the first broken bond according to (f) type of pristine N-carbophene and (h) functionalized 3-carbophene.}
    \label{fig:bootstrap}
\end{figure*}

Figure~\ref{fig:bootstrap} presents examples of the phase-change onset with (a) type 2 bond breaking in graphenylene, (b) a type-2 bond breaking in 3carb2NH2, (c) a type-7 bond breaking in 4-carbophene, and (d) a type-10 bond breaking in 5-carbophene. Each image is a close-up of the broken bond at the time step when it breaks; the bottom-left corner of each image shows the temperature at that time step. Graph (e) in Figure~\ref{fig:bootstrap} presents box-and-whiskers plots of the temperatures at which the first bond breaks in each replicate for each type of N-carbophene and each supercell size. The black I-shapes in each box-and-whiskers plot give the bootstrap-generated bounds for the 95\% confidence interval of the population mean temperature for each N-carbophene type and supercell size~\cite{infer}. Graph (f) is a set of treemaps displaying the percentage of first broken bonds by bond type for the pristine N-carbophenes~\cite{treemapify}. Graph (g) is a box-and-whiskers plot of the temperatures at which the first bonds broke in functionalized 3-carbophene with population mean confidence levels. Finally, Figure~\ref{fig:bootstrap} (h) shows a set of treemaps displaying the percentage of each type of first-broken bond in the functionalized 3-carbophenes.

Two trends emerge in the pristine N-carbophene results in Figure~\ref{fig:bootstrap} (e). The first trend is the effect that supercell size has on the mean onset temperature. Attempts to fit the raw temperature versus the number of C-C bonds and bootstrapped temperature versus the number of C-C bonds both yielded inconclusive results. For example, the raw 3-carbophene temperature values were fit with several models, with both polynomial and power fits giving considerably better fits than other models, yet with R$^2$ values differing by only a few thousandths. But, in the limit as the number of bonds increases to infinity, the polynomial model goes to infinite temperature while the power model goes to zero. Thus, we expect that larger N-carbophene models would need to be created and analyzed to produce a better temperature vs. number of bonds model fit. 
The second trend in Figure~\ref{fig:bootstrap} (e) is that the mean onset temperature decreases with increasing N. Figure~\ref{fig:bootstrap} (f) demonstrates that the cause of the decreasing onset temperature as N increases is due to increasing antiaromaticity near the center of the linear N-phenylene chains that make up each pristine N-carbophene structure (e.g., type 4 bonds for 3-carbophene, type 7 bonds for 4-carbophene, and type 10 bonds for 5-carbophene)~\cite{junkermeier2019simplecarbophene}. Images (g) and (h) in Figure~\ref{fig:bootstrap} display the effect of replacing pristine N-carbophenes with functionalized N-carbophenes.

Chemists interested in probing foundational issues concerning regions of aromatic versus antiaromatic character have long studied linear and angular N-phenylenes~\cite{BARRON19662609}. Algebraic structure count (ASC) analysis of N-phenylenes predicts the following: linear N-phenylenes, where N is the number of six-membered rings, have ASC = N + 1; angular N-phenylenes have ASC equal to the (N + 1)th Fibonacci number; and branched phenylenes have even larger ASC values~\cite{Gutman19932413}. The increasing ASC value signifies greater thermodynamic stability. These ASC results compare favorably with our results. In the N-carbophenes, increasing N changes the nature of the material from graphenylene, which is similar to branched phenylenes, to 5-carbophene, which is more similar to linear phenylenes, leading to lower thermodynamic stability.

Subsequent DFT analysis of linear and angular phenylenes supports the ASC results, demonstrating that linear phenylenes are less stable than angular phenylenes when both contain the same number of six-membered rings~\cite{MaksicLinvsAngPhenylenes}. The increased antiaromatic character of the four-membered rings in linear phenylenes is the likely explanation for this decreased stability. Similarly, in N-carbophenes, bond-length alternation increases in the four-membered rings as N increases, again suggesting a cause for the lowered phase-change temperatures with increasing N~\cite{junkermeier2019simplecarbophene}.

\subsection{Two-Way ANOVA and Tukey's HSD} 

We assessed the size effect in pristine N-carbophene models using a two-way ANOVA to evaluate how categorical variables N and Size influence Temperature. The ANOVA was appropriate given the large group sizes (i.e, $n \ge 30$), despite one of the 12 modalities failing normality tests~\cite{soetewey2023twowayanovainr}. Following ANOVA, Tukey’s Honestly Significant Differences (HSD) post hoc test identified which pairs of modalities differed~\cite{alma9924029750902466, agricolae23}. Table~\ref{table:NCarbHSD.test} reports these results. Our key finding is that for graphenylene and 4-carbophene, mean temperatures differ significantly with size, indicating that size 6 supercells likely underestimate phase-change temperatures. In contrast, for 3-carbophene and 5-carbophene, Size 4 and Size 6 do not differ significantly in mean temperature, suggesting Size 6 is sufficient for reliable predictions. The Tukey HSD results also indicate that the phase-change onsets are not significantly different for 4- and 5-carbophenes. While we may not observe a statistically significant change in the phase-change onset temperatures, we expect it to occur due to Carnelley's rule~\cite{YALKOWSKY2014}. Yang \textit{et al.}'s prediction that 3-carbophene and 5-carbophene are second-order topological insulators further motivates the need for sufficiently large supercells (3carb6P and 5carb6P) to ensure that temperature-induced phase changes are captured accurately~\cite{Yang2025033101}.

\begin{table}[htb]
\centering
\caption{Pristine N-carbophene models with the same letter (i.e., a, b, c, d, e, f, g, h, or i) are not significantly different.}
\begin{small}
\begin{tabular}{rccc}
\toprule

      & Size 2 & Size 4 & Size 6  \\
\midrule
graphenylene & a & b & cd \\
3-carbophene & bc & fg & gh \\
4-carbophene & de & gh  & i \\
5-carbophene & ef & hi & i\\
\bottomrule
\label{table:NCarbHSD.test}
\end{tabular}
\end{small}
\end{table}

\subsection{Negative Area Thermal Expansion}

\begin{figure*}
    \centering
    \includegraphics[clip,width=6.5 in, keepaspectratio]{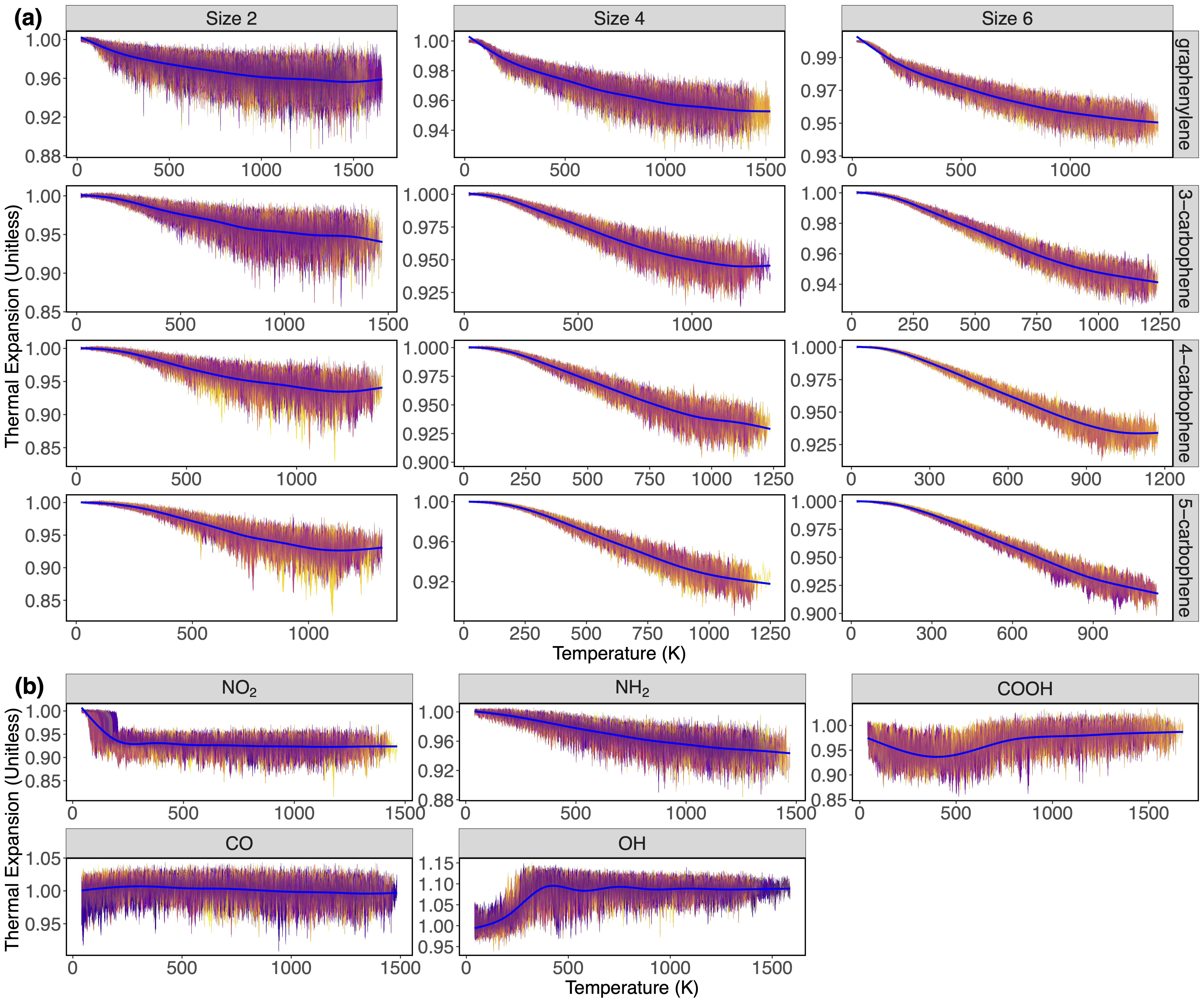}
    \caption{Graphs showing the area thermal expansion versus temperature for each simulation are discussed. The jagged multicolor graph within each facet consists of overlapping plots of area thermal expansion, with each simulation plotted in a different color. The smooth blue line represents the LOWESS smoothing of the data in each facet. (a) Facet plot of the pristine N-carbophenes. (b) Facet plot of the functionalized 3-carbophenes.}
    \label{fig:areavstempAHNPT}
\end{figure*}

In Figure~\ref{fig:areavstempAHNPT} (a), each graph presents a plot of the in-plane thermal area expansion of the periodic bounding box as a function of temperature for each independent replicate of each pristine structure. The plot of each simulation is cut off at the first bond break (the phase-change onset), and the resulting values are presented in Figure~\ref{fig:bootstrap}. The blue line in each graph is the Locally Weighted Scatterplot Smoothing (LOWESS) mean of the area expansion analysis of the 50 replicates. The 95\% confidence interval is within the width of the blue line on the graphs. According to the simulations, pristine N-carbophenes display NATE between 10 K and 1500 K. This trend is consistent with the NATE of graphene between 100 K and 1000 K, which was investigated using MD simulations, quasi-harmonic approximation, and Raman Spectroscopy~\cite{yoon2011negative, Islam2013435302, Magnin2014185401, Mann201722378, Feng2021202006146}.

Replacing a single H atom with a different functional group in a large sheet of pristine N-carbophene creates a defect. In contrast, replacing all of the H atoms of 3carb2P with the same functional group effectively creates a different material. Accordingly, the changes in the form of the area thermal expansion as shown in Figure~\ref{fig:areavstempAHNPT} (b) may be expected. Materials that undergo NATE are technologically valuable in composites and sensors. Thus, thermal-mechanical properties can be tailored by tuning the thermal expansion via chemical functionalization.

\subsection{Functional group stability}

\begin{figure}
    \centering
    \includegraphics[clip,width=\linewidth, keepaspectratio]{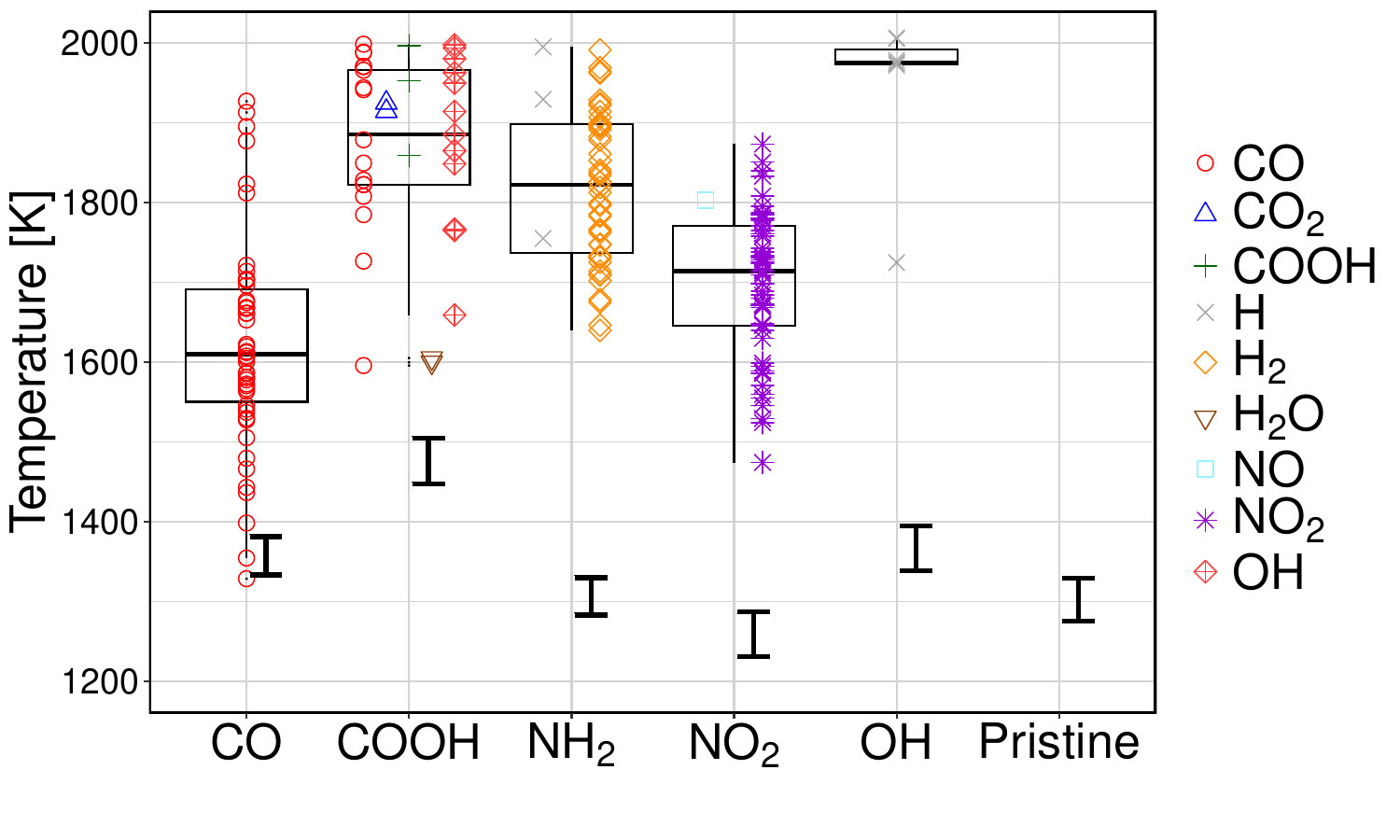}
    \caption{Box-and-whiskers plots of mean desorption temperature of the first atom or molecule to desorb from functionalized 3-carbophenes. Each box-and-whiskers plot shows the molecule that first desorbed and the temperature at which it desorbed.}
    \label{fig:desorbedfunctionalplot}
\end{figure}

Once the phase-change initiates, significant rearrangement occurs, yet the N-carbophenes often hold onto their functional groups until much higher temperatures, as shown in Fig.~\ref{fig:desorbedfunctionalplot}. The absence of many data points for the 3carb2OH and for 3carb2P is due to the majority of N-carbophenes holding onto their functional groups up through 2010 K. Several of the 3carb2COOH also held onto their functional groups above 2010 K. These covalent bond cleavage (desorption) temperature results are consistent with earlier work, which used formation energies to demonstrate that covalent functionalization is stable~\cite{Junkermeier2022Covalent}. 

The thermal stability of the functional groups is important for two reasons. First, Du \textit{et al.} suggested they produced carbophene because their material contained oxygen~\cite{du1740796}. The high desorption temperatures here demonstrate that oxygen-containing groups could survive the synthesis. Second, functionalized N-carbophenes could serve in solid-state gas adsorption, where heating may expel adsorbed gases~\cite{JUNKERMEIER2024112665, JUNKERMEIER2024112921}. Further, these results suggest that N-carbophenes may be candidate materials for functional membranes in high temperature applications.

\subsection{\texorpdfstring{Graphenylene to $\gamma$-graphyne}{graphenylene to gamma-graphyne}}

\begin{figure}
    \centering
    \includegraphics[clip,width=\linewidth, keepaspectratio]{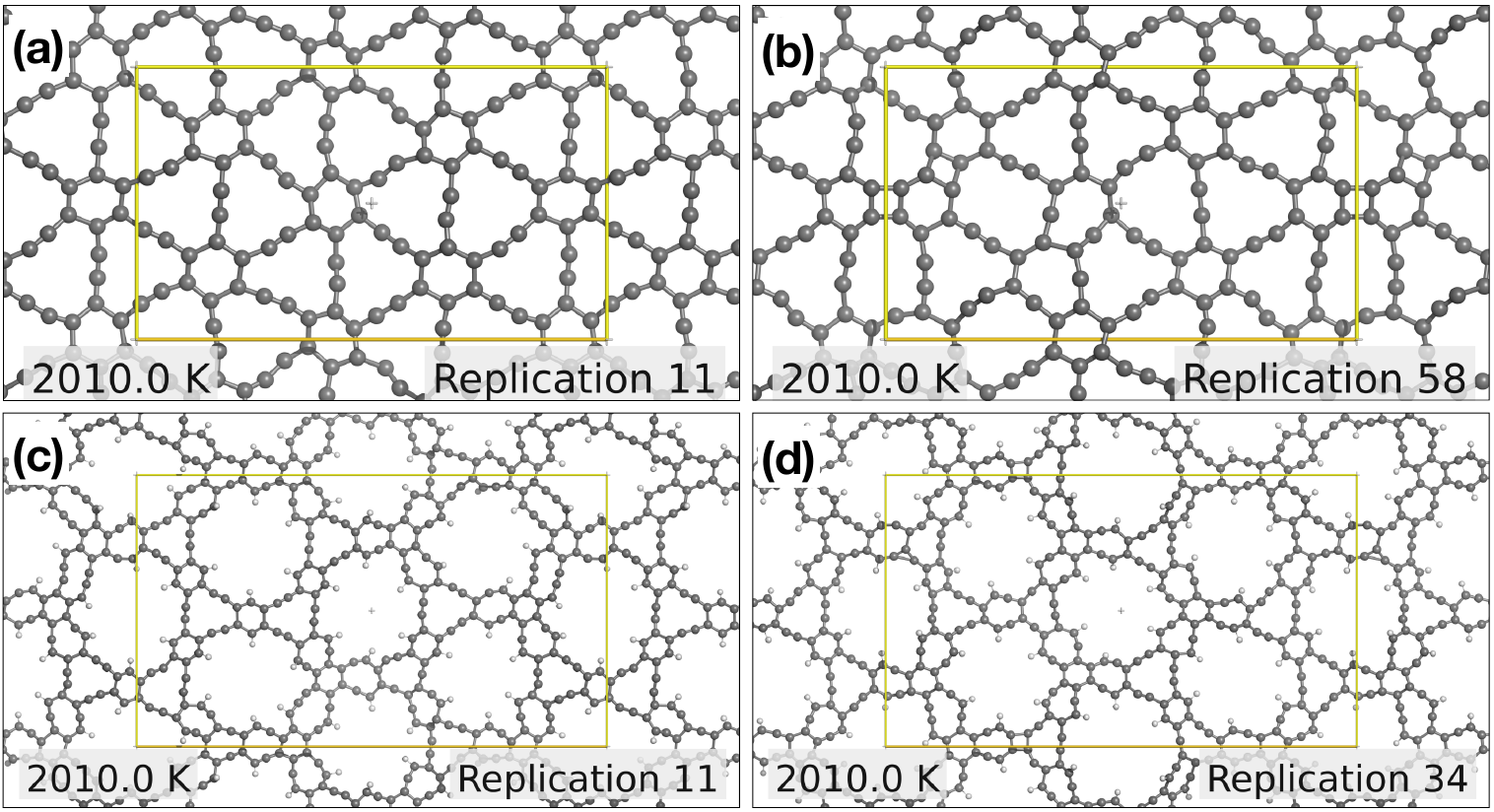}
    \caption{$\gamma$-graphyne (a) and $\gamma$-graphyne with defects (b) produced by heating graphenylene. Images (c) and (d) are the results of 3-carbophene temperature-ramp MD simulations. Gold lines outline supercell boundaries.}
    \label{fig:graphenylene2gammagraphyne}
\end{figure}

The MD simulations suggest that graphenylene and 3-carbophene may undergo phase transitions to $\gamma$-graphyne and a porous $\gamma$-graphyne structure, respectively. 
Thirteen out of the fifty 2carb2P replicates transitioned into $\gamma$-graphyne structures by 2010 K, an example is given in Figure~\ref{fig:graphenylene2gammagraphyne} (a).  
The rest of the 2carb2P replicates appeared to transition into $\gamma$-graphyne with defects, sometimes with a phenylene-like pattern still in it, see Figure~\ref{fig:graphenylene2gammagraphyne} (b). None of the 2carb4P or 2carb6P replicates transitioned into $\gamma$-graphyne, though all of these appear to be similar to $\gamma$-graphyne with many defects. Similarly, the 3-carbophene models all appear to be moving towards a type of porous $\gamma$-graphyne; though none of the replicates do so cleanly. To evaluate the thermodynamic viability of the phase changes, formation energy calculations were performed for all relevant phases. Full details of the computational methods, reference states, and formation energy results are provided in the Supplemental Information.

While $\gamma$-graphyne may be synthesized experimentally, this result suggests that graphenylene may be a synthetic precursor material for graphyne-like structures at high temperatures~\cite{LI2018248}. The fact that only small supercells (2carb2P) fully transition while larger ones do not raises interesting questions about finite-size effects, defect propagation, and domain cooperative behavior.  This result also raises questions about whether the recently synthesized biphenylene-based material net-C could be annealed into another carbon allotrope~\cite{tyutyulkov1997structure, Fan2021852}.

\section{Conclusion}
This study aimed to determine the thermal stability thresholds and high-temperature response of pristine and functionalized N-carbophenes using RMD simulations. Phase-change onset temperatures decrease systematically with increasing N-phenylene chain length, reflecting the increasing antiaromaticity of the central phenylene segments; however, they remain above 1000 K in all cases. While pristine N-carbophenes exhibit NATE, we can tune the nature of the thermal expansion by selectively functionalizing the N-carbophenes to achieve either positive or negative thermal expansion. Functional groups (CO, COOH, NH$_2$, NO$_2$, OH) remain covalently bound well above the onset of structural transformation. A temperature-induced graphenylene to $\gamma$-graphyne transformation reveals a previously unrecognized high-temperature synthesis pathway. More broadly, these results demonstrate that biphenylene-based 2D carbons are stable at temperatures required for synthesis and device operation~\cite{du1740796}. Further, functionalization is a viable route to tuning thermal and mechanical behavior without compromising chemical stability. Future work could focus on expanding to larger supercells and incorporating additional functional groups to elucidate further the phase-change temperatures and thermal expansion properties of N-carbophenes, as well as the reversibility of structural transformations and aromaticity-driven structural weaknesses. Further, work should also be conducted concerning defect propagation and domain cooperative behavior in the graphenylene to $\gamma$-graphyne transition. N-carbophenes are robust, designable 2D carbon materials with promise beyond graphene

\section{Acknowledgements}

C. Junkermeier was supported by the United States funding agency NSF under Award \#2113011, an ReaxFF Unlimited seat Group License from Software for Chemistry \& Materials (SCM) provided by SCM, Inc., and advanced computing resources from University of Hawaii Information Technology Services – Cyberinfrastructure, funded in part by the National Science Foundation awards \#2201428 and \#2232862. K. Lavarez, R. Adra, and V. Diaz were supported in-part with funding from the Undergraduate Research Opportunities Program in the Office of the Vice Provost for Research and Scholarship at the University of Hawai‘i at Mānoa (Project ID: 80847-ASTR). H. Osterstock was supported by STEMworks Hawaii internship, a program of the Maui Economic Development Board, Inc. R. Paupitz acknowledges Brazilian funding agencies: FAPESP (grant \#2021/14977-2) and CNPq (grant \#313592/2023-3). 

This research was conducted with the assistance of students from Kihei Charter School, King Kakaulike High School, and Maui High School, as well as at Discover UH Manoa Open Houses.

\section{{CReDiT} authorship contribution statement}

Chad E. Junkermeier: Conceptualization, Data curation, Formal analysis, Investigation, Methodology, Project administration, Resources, Software, Validation, Visualization, Supervision, Writing – original draft, Funding acquisition.
Kat Lavarez: Investigation, Formal analysis, Writing – original draft, Funding acquisition
R. Martin Adra: Investigation, Formal analysis, Writing – original draft, Funding acquisition
Valeria Aparicio Diaz: Investigation, Formal analysis, Writing – original draft, Funding acquisition
Heather Osterstock: Investigation, Formal analysis, Writing – review \& editing.
Pal Casinto: Investigation, Writing – review \& editing
Meghan Worrell: Investigation, Writing – review \& editing
Ricardo Paupitz: Writing – review \& editing.
Adri C. T. van Duin: Resources, Software, Writing – review \& editing.

\section{Declaration of generative {AI} and {AI}-assisted technologies in the manuscript preparation process}
During the preparation of this work the author(s) used ChatGPT in order to develop the Abstract and Conclusion section. After using this tool/service, the authors reviewed and edited the content as needed and take(s) full responsibility for the content of the published article.



\bibliographystyle{elsarticle-num.bst}
\bibliography{carbophene}





\end{document}